\documentclass[10pt]{IEEEtran}
\usepackage[left=1.62cm,right=1.62cm,top=1.9cm]{geometry}
\usepackage{cite}
\usepackage{amsmath,amssymb,amsfonts}
\usepackage{algorithm}
\usepackage{algpseudocode}
\usepackage{graphicx}
\usepackage{textcomp}
\usepackage{xcolor}
\usepackage{caption}
\usepackage{multirow}
\usepackage[normalem]{ulem}
\usepackage[T1]{fontenc}
\usepackage{svg}

\def\BibTeX{{\rm B\kern-.05em{\sc i\kern-.025em b}\kern-.08em
    T\kern-.1667em\lower.7ex\hbox{E}\kern-.125emX}}
\usepackage{url}

\begin{document}

\title{
\textcolor{black}{Data-driven Software-based Power Estimation for Embedded Devices}
%\textcolor{red}{Software-Level Power Estimation for Embedded Devices through Machine Learning %and %Profiling}
}

\author{
Haoyu Wang \IEEEauthorrefmark{1}, 
Xinyi Li\IEEEauthorrefmark{2}, Ti Zhou\IEEEauthorrefmark{3}  and
Man Lin\IEEEauthorrefmark{4}
\thanks{Department of Computer Science,
St. Francis Xavier University,
Canada. 
Email: 
\IEEEauthorrefmark{1}x2020fcw@stfx.ca,
\IEEEauthorrefmark{2}x2021gim@stfx.ca,
\IEEEauthorrefmark{3}tizhou1@cs.stonybrook.edu,
\IEEEauthorrefmark{4}mlin@stfx.ca.}
\thanks{
A version appears at
archive: https://arxiv.org/abs/2407.02764. 

The research is supported by Natural Sciences and Engineering Research Council of Canada.}
}

\maketitle
\begin{abstract}
\textcolor{black}{
Energy measurement of computer devices, which are widely used in the Internet of Things (IoT),  
is
an important yet challenging task. 
Most of these IoT devices lack ready-to-use hardware or software for power measurement.  
In this paper,  we propose an easy-to-use approach to derive a software-based energy estimation model with external
low-end power meters  based on data-driven analysis.
 Our solution is demonstrated
with a Jetson Nano board and Ruideng UM25C USB power
meter. Various  machine learning methods combined with our smart data
collection \& profiling method and physical measurement are explored. Periodic Long-duration measurements are utilized in the experiments to derive and validate power models, allowing more accurate power readings from the low-end power meter.
Benchmarks 
were used to evaluate the derived software-power model for the Jetson Nano board and Raspberry Pi. The results show that  92\% accuracy can be achieved by the software-based power estimation compared to measurement.
A kernel module that can collect running traces of utilization and frequencies needed is developed, together with the power model derived, for power prediction for programs running in a real environment. 
Our cost-effective method facilitates accurate instantaneous power estimation, which low-end power meters cannot directly provide.
}

\end{abstract}


\section{Introduction}

Measuring the energy consumption of computer systems has always been an important task, as:
\begin{itemize}
    \item Energy consumption is an important indicator of system performance \cite{weiser1996scheduling} \cite{baynes2003performance} \cite{sadi2017joint}. In the realms of head-mounted AR \cite{choi2017analyzing} and edge computing \cite{shi2016edgecomputing}, 
    energy efficiency directly influences operational capabilities and sustainability.
    \item Some energy-aware system design methodology makes control based on energy consumption feedback \cite{schurgers2003power} \cite{chen2022energy}.
    
 \end{itemize}

High-end and accurate hardware-based measurement \cite{carroll2010analysis} \cite{raghunathan2001adaptive} \cite{xu2004practical} is one solution,  but such hardware is often expensive and requires specialized operation and integration knowledge. To allow users to measure the energy
consumption of their devices without the need of specialized hardware, vendors offer several software-level solutions, such as Intel RAPL \cite{david2010rapl} and AMD uProf \cite{AMDuprof}.
Such tools estimate energy consumption through statistics collected from runtime hardware devices.
Building a software-level energy analysis tool is often accompanied by building an energy model \cite{zhang2010accurate} \cite{kumar2022machine} \cite{arda2020ds3} \cite{djedidi2020power} and ensuring that the hardware devices can provide the parameters needed for that model with low overhead.

Unfortunately, as of today, many popular embedded hardware 
that is widely used in IoT and Mobile Edge Computing (MEC) System,   lacks a software-level approach for energy measurement.
In this case, users who do not have access to high-end equipment can only measure power consumption based on consumer-grade meters \cite{liu2021dart} 
\cite{zhou2023profiling}.
We next discuss the limitations of using consumer-grade meters.

\subsection{Limitations of using consumer-grade power meters for energy measurement}

\textbf{Consumer-grade power meters often do not provide data API.}
To integrate power measurement into energy-saving policy design, a practical solution will be using the data interface provided by the meter (similar to the \textit{sysfs} interface in Linux) for programs to access power meter measurement.
Unfortunately, we did not find a consumer-grade energy meter that offers a data API. In most cases, the power meter provides a GUI interface for the user to read data from the hardware or software connected to the meter via Bluetooth \cite{um25cdoc}.
The lack of a low-overhead and low-latency data access scheme means that designing low-power management methods based on energy feedback for control is challenging with consumer-grade power meters.

\begin{figure}[h]
  \centering
\includegraphics[width=0.8\linewidth]{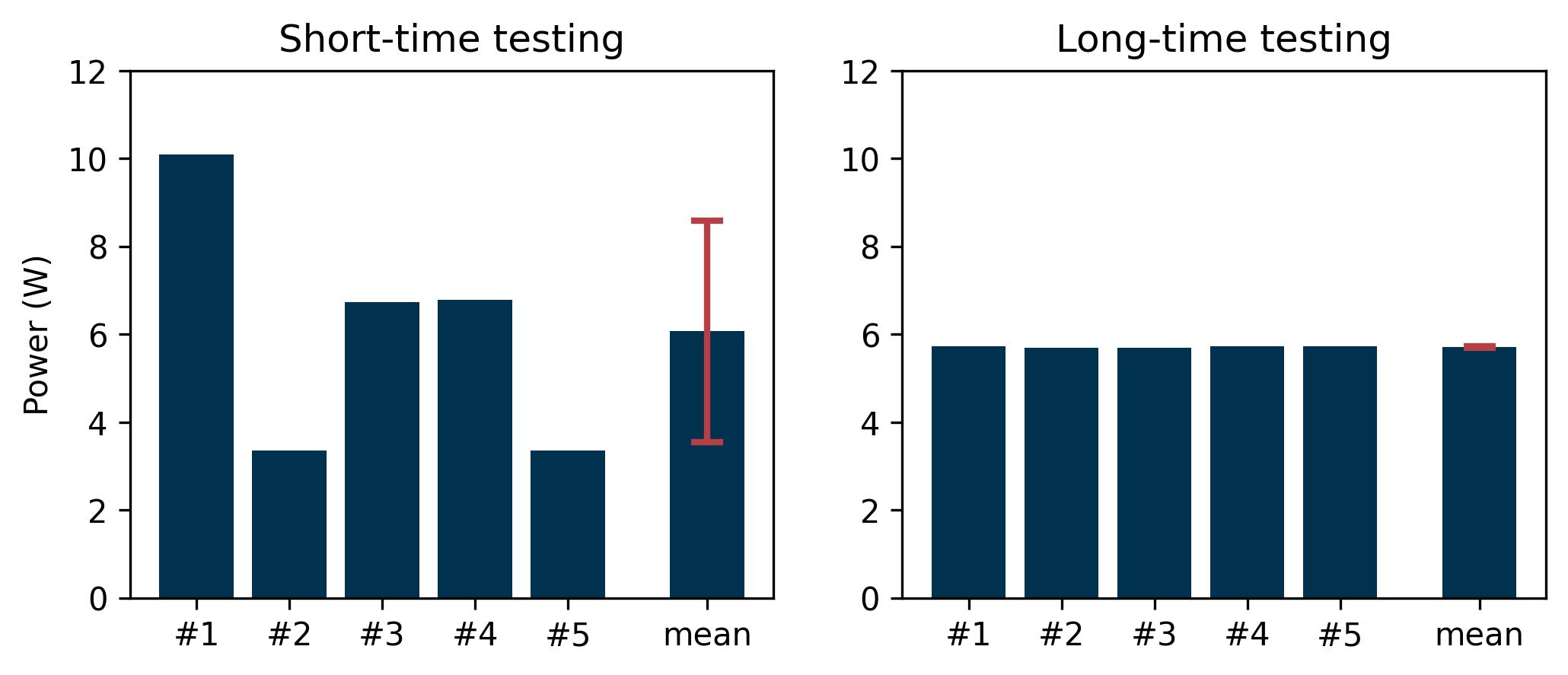}
  \caption{Power measuring using a consumer-grade meter.}
  \label{fig:power}
\end{figure}

\textbf{Consumer-grade power meters cannot provide accurate short-duration measurement.} 

Due to hardware limitations, consumer-grade meters record voltage/current changes in a coarse-grained view, usually in seconds. This means it is difficult to accurately measure energy consumption over short periods of time. 
Next we illustrate this point using an example of measuring the energy consumption of a CPU-intensive program with an execution time of around 1 second.
We applied two methods. One is based on short-duration measurement, where we simply record the energy consumption during the 1-second execution. Another one is based on long-duration measurement, where we run the workload repetitively in a loop for three minutes and obtain the average energy consumption for 1-second execution as the total 3-minute energy consumption divided by the number of executions.
The test is shown in Fig. \ref{fig:power}. The variance of measurements over short periods of time is much larger than over long periods of time.
This illustrates that short-duration measurement is not reliable and only measurements over a long period of time can provide an accurate, consistent result. 

Because of the above-discussed limitations, energy measurements in our previous works \cite{zhou2022deadline} \cite{zhou2023cpu}, were very labor intensive. The work proposed in this paper overcomes the above limitations.

\subsection{\textcolor{black}{Proposed approach: data-driven power estimation }}
\textcolor{black}{
In this paper,
we propose a data-driven method to derive software-based power estimate models using affordable external consumer-grade meters,  and provide a profiling tool to estimate the power of  programs.
The contributions of the paper can be summarized as follows.
\begin{itemize}
    \item \textcolor{black}{We propose a method to construct a software-based energy consumption estimation model for  small devices.}
    \item Our method is based on smart data collection and data analytics techniques to estimate power. Our data collection method is simple yet able to sample the parameter space evenly. 
    \item \textcolor{black}{
    We also provide a platform-independent kernel module that can be used to collect running traces of utilization and frequencies needed for power prediction for programs running in Linux environment.} 
    \item \textcolor{black}{The approach can help users estimate energy on devices lacking integrated energy measurement software, especially when they only have access to affordable consumer-grade meters.}
    \textcolor{black} {Our approach takes into account  short-duration estimation error, and overcomes the limitation of the consumer-grade power meters.}
    \item Our energy estimation model can be used to obtain timely feedback on energy consumption, which is especially useful for AI researchers to design or improve energy control policies as it avoids the inconvenience of integrating hardware-based measurements.
\end{itemize}
}

\subsection{\textcolor{black}{Related works}}

\textcolor{black}{Edge computing on small devices has become a crucial component of today’s intelligent society and has attracted significant research interest \cite{sadi2017joint} \cite{shi2016edgecomputing} %\cite{chen2022energy} update
%\cite{zhou2019security} 
\cite{kumar2022machine} \cite{djedidi2020power} \cite{liu2021dart} %\cite{alsahli2021privacy}
\cite{zhou2023cpu} \cite{zhou2023profiling} \cite{zhou2022deadline} \cite{reghenzani2021multi} %\cite{ranjbar2020power} %\cite{bhuiyan2018energy} %\cite{bhuiyan2020energy} 
%\cite{wu2023devfuzz}
\cite{wu2024intos} 
%\cite{guo2019energy} 
%\cite{huang2022energy} %\cite{bakshi2019fast} 
\cite{ayvaz2021predictive} \cite{panda2022energy} \cite{hoffmann2021online} %\cite{park2021interpretable} 
\cite{wang2021online}.}
\textcolor{black}{In system design, unlike design solutions that are based on mathematical models for offline analysis\cite{reghenzani2021multi} %\cite{ranjbar2020power} 
%\cite{zhu2003scheduling}  
\cite{bhuiyan2018energy}%\cite{bhuiyan2020energy} %\cite{guo2019energy}
%\cite{huang2022energy} 
\textcolor{black}{\cite{zhou2023energy}}, adaptive system design and measurement methods \cite{zhou2022deadline} \cite{zhou2023cpu} %\cite{bakshi2019fast} 
\cite{ayvaz2021predictive} \cite{panda2022energy} \cite{hoffmann2021online} %\cite{park2021interpretable}
\cite{zhang2023infinistore} 
\cite{ma2024malletrain} \cite{ali2025enabling}
\cite{das2015workload} \cite{wang2016model} \cite{wang2021online} %\cite{shafik2015learning} 
\cite{DeepakUic2021} 
\textcolor{black}{\cite{zhao2024efficient}  \cite{han2024kace} \cite{li2024energy}}
have received attention in recent years.}
Accurate power modeling and measurement have been widely explored in the literature, with various approaches proposed for different types of processors and systems. 
Below, we review some key contributions and compare them with our proposed method.

\textcolor{black}{Rethinagiri et al. \cite{rethinagiri2014system} built power models for ARM-based processors, such as Cortex-A8 and Cortex-A9. They measure static and dynamic currents with an Agilent LXI digitizer and use a multimeter to measure the voltage on the jumpers to get the power. The parameters used to build the model are frequency, instructions per cycle (IPC), cache miss rate, etc. The models are built differently for different processors. They built accurate models for the processors and used measuring instruments with high precision. In contrast, our work proposes a method that allows researchers who can only use consumer-level devices to construct energy consumption models for their own devices. Also, the parameters we chose are common to all processors, while their work chose parameters based on the specific processor, including some low-level parameters.}

\textcolor{black}{Walker et al. \cite{walker2016accurate} also derived models for ARM-based processors. Their approach is noteworthy because they collected data from multiple PMCs and then constructed algorithms to select the PMC parameters that were ultimately put into the model. Nikov et al. \cite{nikov2022accurate} constructed a linear regression model for Cortex-M0 using a total of 6 parameters such as executed instructions, multiplication instructions, taken branches and RAM data reads. They measure energy consumption using a custom measurement board.}

\textcolor{black}{There are some methods to build a power model from the RTL level. Zoni et al. \cite{zoni2018powerprobe} proposed a method that could implement a power model based on the RTL description of the target architecture. Kim et al. \cite{kim2019simmani} collected 50 signals at the RTL level and then proposed a method to select the signals to build the model automatically. These methods are characterized by a more complex model-building process and require domain-specific knowledge. In contrast, our proposed approach is easy to follow, reproducible, and does not require much domain-specific knowledge of CPU architecture.}

\textcolor{black}{Some works also build power models, but they do not use real machine measurements. They use simulated power \cite{xie2021apollo} \cite{zhan2021deepmtl} \cite{zhan2022deepmtl} or simulators \cite{li2005power}. Simulated models provide insights and theoretical frameworks but often lack the accuracy and real-world applicability of models built with actual hardware measurements. Our approach is performed on real devices for energy measurement as well as model testing, ensuring practical applicability and accuracy.}

\textcolor{black}{Table~\ref{table:relatedWParameters} and Table~\ref{table:Measurement} summarize the key parameters used in the power estimation models and energy acquisition  methods for different approaches. Note that PMC stands for Performance Monitoring Counter.}
\begin{table}[htbp]
\centering
\caption{\textcolor{black}{The Key Parameters for Power Estimation}}
\label{table:relatedWParameters}
\begin{tabular}{|l|l|}
\hline
\textbf{Types} & 
\textbf{Works} \\\hline
Register Transfer Level (RTL) Model & \cite{zoni2018powerprobe}, \cite{kim2019simmani}  \\\hline
Micro-Architecture Model + Code Trace
& \cite{McPAT}~+~\cite{Germ5} \\\hline
Instruction Level Energy Model
& \cite{InstructionLevel96}  \\\hline
PMC: Executed instructions, branch ... & \cite{nikov2022accurate} \\\hline
PMC: Instruction per Cycle, Cache Miss Rate, ...
& \cite{rethinagiri2014system} \\\hline
PMC: Select events for processor-specific configuration & \cite{walker2016accurate} \\\hline
Frequency, Utilization & Our Work\\
\hline
\end{tabular}
\end{table}

\begin{table}[htbp]
\centering
\caption{\textcolor{black}{Energy Acquisition Methods }}
%used for Validation}}
\label{table:Measurement}
\begin{tabular}{|l|l|}
\hline
\textbf{Types} & 
\textbf{Works} \\ \hline
High Precision Instrument & \cite{rethinagiri2014system} \\ \hline
Custom Board & 
\cite{InstructionLevel96}
\cite{nikov2022accurate}  \\\hline
Build-in Energy Sensor & 
\cite{walker2016accurate}
  \\\hline
Consumer Grade Meter & Our Work\\\hline
Simulator &  
\cite{xie2021apollo} \cite{zhan2021deepmtl} \cite{zhan2022deepmtl}
\cite{li2005power} \\
\hline
\end{tabular}
\end{table}

\textcolor{black}{In summary, while existing methods provide high accuracy and detailed models, they often involve high costs, complexity, and specialized knowledge. Our approach offers a practical alternative that balances ease of use, cost-effectiveness, and accuracy, making it suitable for embedded and IoT devices lacking built-in power measurement capabilities.}

\section{Data-driven Power Estimation with Consumer-grade Meters}
Our data-driven method to derive a software-based power estimate model using only consumer-grade meters is divided into three main steps.
\begin{enumerate}
    \item Collect data related to workload execution and energy using long-duration energy measurement with power-meter.
    \item Construct energy model based on collected data from the devices and the power meter.
    \item Estimate the power of the workloads using the model constructed and a software module that can track workload executions.
\end{enumerate}

\begin{figure}[h]
  \centering
  \includegraphics[width=\linewidth]{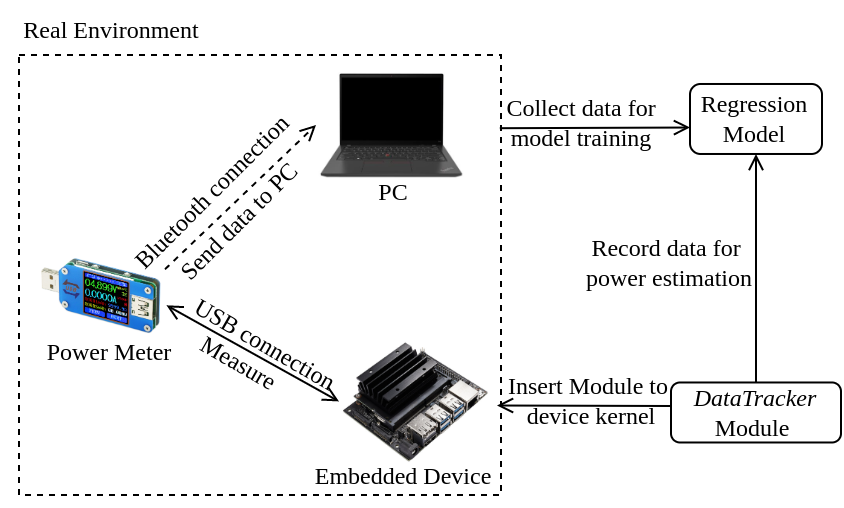}
  \caption{An illustration of the methodology}
  \label{fig:method}
\end{figure}

The hardware connection for the approach is shown  in Figure~\ref{fig:method}. The power meter is connected to the underlying device to be studied. Note that workload will run on the device for data collection to construct energy model for the device. A PC is used to receive the Bluetooth data sent by the power meter, which is energy consumption data. In addition, the parameters (frequency and utilization of the device) that are used to build the model, are recorded when the workload runs on the device. These data are integrated with the energy consumption data for energy-model construction. A data collection module is required to retrieve the necessary parameters like the frequency and utilization data for using the model.

\subsection{ Training Data  Collection}
\label{ch:traindata}
The way to collect training data is to run benchmark programs and measure the energy consumption with a power meter while collecting corresponding parameter traces while the benchmark runs.
There are many factors that can affect power, such as voltage, current, frequency, utilization, etc. Among these parameters, we choose frequency and utilization because there is a strong correlation between these two parameters and the power.  Utilization is an empirical choice and there are other studies \cite{walker2016accurate} that consider utilization when constructing power models or energy models. Based on the equations of the CMOS circuit model \cite{weste2004cmos}, frequency and power show a sub-square relationship, which means frequency and power are strongly correlated.

In order to collect observations of evenly distributed utilization and frequency levels, we choose CPU-intensive workloads as our benchmarks so that we can control the frequency and utilization easily.
 One of the benchmarks is a single-threaded benchmark, which performs multiple multiplications. This benchmark's utilization on a single core is 100\%. The embedded device that we perform our experiments is a  Nvidia Jetson Nano Board 2GB. The number of cores on the board is 4. Thus, the average utilization for the four cores is 25\%, which means the highest utilization for running this benchmark is 25\%. The other benchmark is a multi-threaded benchmark, which runs multiple multiplication programs, each of which is 100\% CPU-intensive on one core. In our experiments, there are 4 threads in total.

Since the data to be collected are related to frequency and utilization, we controlled (regulated) both frequency and utilization levels (evenly distributed) to different values for data collection, which is shown in Algorithm~\ref{algo:Measurement}. In this algorithm, the frequency is controlled by setting the DVFS governor to $userspace$ and then running the benchmark at each frequency.  Controlling utilization levels is relatively more complicated. Different utilization levels are obtained by setting a proper combination of runtime and idle time. For example, an approximate utilization of 50\% can be obtained by first running the benchmark and then letting the device stay idle for the same amount of time. \textcolor{black}{The purpose of regulating the frequencies to all available levels and utilization to evenly spaced values is to allow simple data sampling with good data distribution.}

The total runtime of each workload is set to 180 seconds. The runtime is controlled by a $factor$, ranging from 0 to 10. By multiplying a different $factor$, the runtime of the benchmark will be different. The idle time is obtained by subtracting the run time from the total time.

In the algorithm, $read\_energy()$ is done through power meter measurements. The utilization is retrieved from the $/proc$ file system.

\begin{algorithm}
%\caption{Measure power and record frequency and utilization}
\caption{Power Measurement for Regulated  Frequency and Utilization Levels}
\label{algo:Measurement}
\begin{algorithmic}[1]
    \State set $total\_time$ to 180 seconds
    \State set $runtime\_factor$ to 18
    \State set governer to userspace
    \For{each $frequency$ that the device support}
        \State set $frequency$
        \For{$factor$ from 0 to 10}
            \State $energy\_start$ $\gets$ read\_energy()
            \State $time\_start$ $\gets$ time.time()
            \While{$time.time() - time\_start < factor \times runtime\_factor$}
                \State run the benchmark
            \EndWhile
            \State $time.sleep(total\_time - $
            \Statex $ \qquad \qquad \qquad factor \times runtime\_factor)$
            \State $energy\_end$ $\gets$ read\_energy()
            \State $time\_end$ $\gets$ time.time()
            \State $power \gets \frac{energy\_end - energy\_start}{time\_end - time\_start}$
            \State $utilization$ $\gets$ system recorder
            \State record ($power$, $frequency$, $utilization$)
        \EndFor
    \EndFor
\end{algorithmic}
\end{algorithm}

\subsection{Derive Power Model from Collected Data}
\label{ch:model}
After obtaining the training data, various methods can be  explored to derive a model to predict the power for a given utilization and frequency. The models that we explored  include regression models,  decision tree models, and neural network models. Note that the types of models to be used are not limited to the above. The details of deriving models for a Jeston Nano board will be described in section~\ref{Sec:ConstructModel}. Here we show the regression model as an example, where $util$ denotes utilization and $freq$ denotes frequency.
\begin{equation}
    Power =  k_0 \cdot util^2 + k_1 \cdot freq \cdot util + k_2 \cdot util + b
\label{eqn:model}
\end{equation}

Different regression models were explored for the Jeston nano board.
Among them, the best results were obtained from the per-frequency regression model.
In this  model, at each frequency the device supports, there is a polynomial expression consisting of utilization, as shown in Equation~\ref{eqn:model}. At each frequency $i$, the corresponding coefficients are $k_0^i$, $k_1^i$, $k_2^i$ and $b^i$. The reason for choosing such a regression model is that the change in frequency could cause a jump in power. More details can be found in Section~\ref{ch:regression}.

\subsection{How to Use the Software-based Power Model?}
\label{ch:usemodel}
Our model is constructed based on benchmark programs measured at a single frequency. However, in a real environment, programs do not often run at a fixed frequency because Operating Systems may dynamically change the frequencies based on system demand. Our model can also predict power for such running scenarios, but the operation becomes relatively complex. 

To predict the power of a program running with multiple
frequencies, it is necessary to obtain all the frequencies
used  running the program, the utilization for each frequency, and
the duration of the frequency. With the utilization of each frequency segment, the power over this frequency period can be obtained based on our model.

The total power of the program can then be calculated by Equation~\ref{eqn:multifreq}, where $n$ represents the total number of frequency segments, $p_i$ represents the predicted power of the segment which is calculated based on the constructed model, and $t_i$ represents the duration of the segment.

\begin{equation}
    \label{eqn:multifreq}
    Power = \frac{\sum_{i=1}^n p_i \times t_i}{\sum_{i=1}^n t_i}
\end{equation}

To get the duration and utilization information for running traces, we built a kernel module called {\it DataTracker} module that can directly track all the frequencies used together with their duration and the utilization of each core at each frequency change.  The detailed information of this module will be explained in Section~\ref{ch:module}.

\section{ Building Power Model}
\label{Sec:ConstructModel}
In this section, we illustrate our method by showing \textcolor{black}{how to construct power models based on regulated data collection. We use a Jetson Nano Board (2GB) as an example. } 

\subsection{Training Data Collection}
The data collection includes setting frequency and utilization for a given workload and measuring the energy consumption of running the workload. We choose frequency and utilization as parameters as they are the two most important parameters that affect CPU energy consumption during a period of time when running a workload under a DVFS policy. Note that CPU energy consumption consumes the vast majority of energy among all the components of the device, especially when there is no display.  Therefore, the model we build will be a good approximation of the energy model device, even though we do not take the detailed model of the CMOS circuit and board architecture as input. 

\subsubsection{Time for Running Workload}
The total run time for each workload is set to 3 minutes. The duration for measurement is set to such a consistently long time because the power meter cannot accurately measure the energy consumption of workloads that run for a short period. The longer the test time, the more accurate it will be. Such a setting guarantees that each measurement of collected energy consumption is sufficiently accurate.

\subsubsection{Frequency Control}
Controlling the frequencies can be achieved by setting the frequency of the workload with the User Governor provided by Linux.
The workload will be running with each available frequencies. 

 \subsubsection{Utilization Control}
To collect data that covers the parameter space evenly, as discussed in Section~\ref{ch:traindata}, we divide the utilization into several levels with equal distance in between and measure the energy consumption for each combination of utilization and frequency.
This method avoids the need to use multiple benchmarks to collect data with different utilization and is more convenient. But note that the level to which we want to control the CPU utilization can only be approximate. For example, if we want to get a 10\% utilization by controlling the running pattern of workload, the real CPU utilization could be 8\%. Thus, we use the target utilization to control the workload and measure the resulting exact utilization for each controlled workload. The utilization data collected for training is the exact utilization for the controlled workload. How to control the utilization of a workload was described by Algorithm~\ref{algo:Measurement} in Section~\ref{ch:traindata}.  

\subsubsection{Training Data Illustration}
\label{ch:resultsoftraindata}
\begin{figure}[h]
  \centering
  \includegraphics[width=\linewidth]{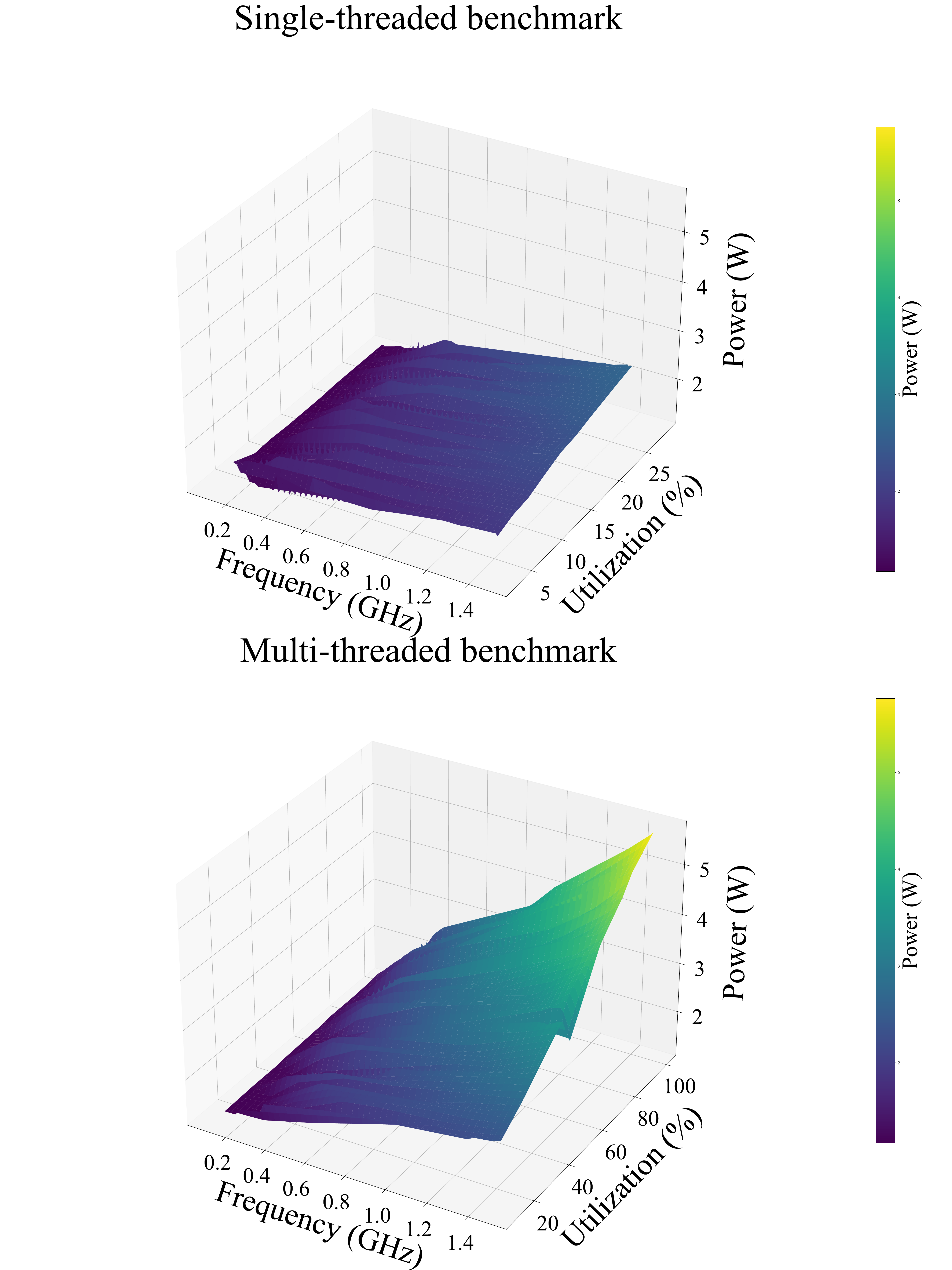}
  \caption{Measurement results of single-threaded (top) and multi-threaded (bottom) benchmark}
  \label{fig:cpu-intensive}
\end{figure}

The embedded device, Nvidia Jetson Nano Board 2GB (JTN), supports a range of frequencies from 0.102 GHz to 1.479 GHz. \textcolor{black}{The supported utilisation range is 0\% to 100\%.} The number of CPU cores is 4. Training data collection is done based on the algorithm explained in Section~\ref{ch:traindata}. The training data
are shown in Fig.~\ref{fig:cpu-intensive}. This is a 3D figure with the x, y, and z axes representing frequency, utilization, and power, respectively.

It can be seen that as the frequency and utilization increase, so does the power. When the frequency is constant, the utilization and power are smoothly related. When the utilization is roughly within a range, there is a jump in power as the frequency changes. This is obvious in the single-threaded benchmark. The first three frequencies correspond to a power level, and the latter frequency corresponds to another power level. We inferred that the reason for this jump is the change in voltage. Comparing the single-threaded data with the multi-threaded data, we can see that the multi-threaded data have a steeper trend. 

We only used multi-threaded data and not single-threaded data when constructing the model as they span across a wide range of utilization.

\subsection{Testing Data Collection}
\label{ch:testdata}

 We chose to select benchmarks (see Table \ref{table:benchmarks} for details) from Mibench \cite{guthaus2001mibench} and Sysbench \cite{kopytov2004Sysbench} as the testing data. These two test sets were chosen because they are both commonly used, lightweight and easy to measure. Mibench contains different types of benchmarks, most of them are single-threaded, while the benchmarks in Sysbench are multi-threaded.
The 10 benchmarks (Table \ref{table:benchmarks}) selected from Mibench cover all the types in Mibench, aiming to mitigate bias.

The benchmarks selected from Sysbench are $cpu$, $memory$ and $fileio$, where $cpu$ performs a series of CPU-intensive computational tasks, $memory$ performs a series of memory-related operations and $fileio$ performs file read and write operations. Since Sysbench is a multi-threaded benchmark test, the number of threads can be set to different values. In the experiment, we set it to 1, 2, 3, and 4 to get multi-threaded data.

\begin{table}[htbp]
\centering
\caption{\textcolor{black}{List of testing benchmark programs}}
\label{table:benchmarks}
\begin{tabular}{|l|l|l|}
\hline
\textbf{Suite} & \textbf{Benchmark} & \textbf{Description} \\
\hline
\multirow{10}{*}{Mibench} & basicmath & Basic mathematical calculations \\
& bitcount & Bit manipulation and counting \\
& jpeg & JPEG image compression and decompression \\
& typeset & Typesetting operations \\
& dijkstra & Dijkstra's shortest path algorithm \\
& patricia & Patricia trie data structure algorithm \\
& stringsearch & String searching algorithms \\
& blowfish & Blowfish encryption algorithm \\
& sha & Secure Hash Algorithm (SHA) computation \\
& crc32 & CRC32 checksum computation \\
\hline
\multirow{3}{*}{Sysbench} & cpu & CPU-intensive computational tasks \\
& memory & Memory-related operations \\
& fileio & File read and write operations \\
\hline
\end{tabular}
\end{table}

\begin{figure}[h]
  \centering
  \includegraphics[width=0.9\linewidth]{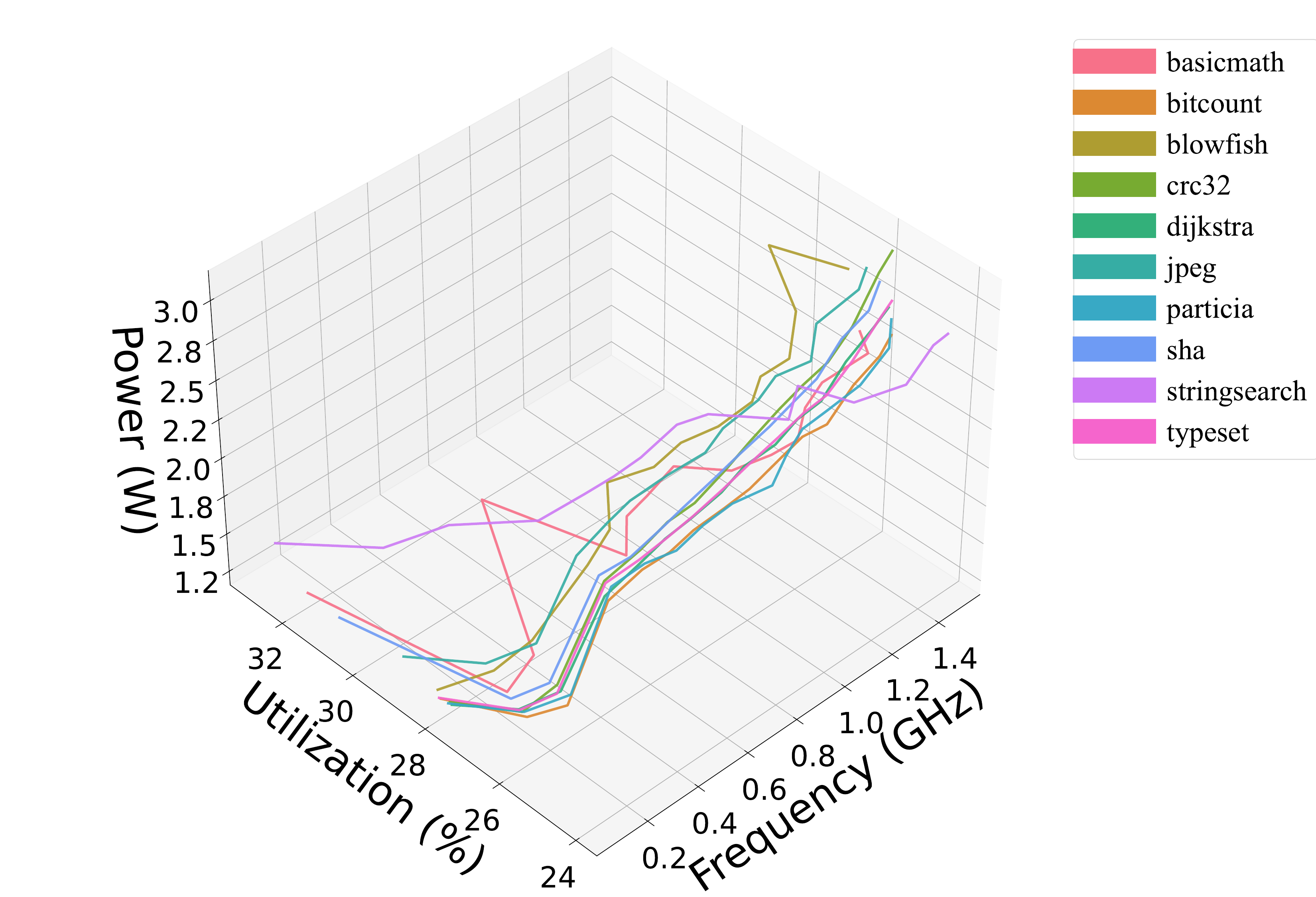}
  \caption{Measurement results of 10 benchmarks in Mibench}
  \label{fig:mibench-9}
\end{figure}

We set the CPU frequency to each available level when running each  Mibench (or Sysbench) benchmark. We then obtain the utilization and the measured power for each running case. The utilization is obtained by reading from the $/proc$ file system, and the power is obtained by a long-time measurement with the power meter. At each frequency, each benchmark is run multiple times for 3 minutes. The purpose of this is to reduce errors introduced by single or short-time measurements. The method of reading energy consumption is the same as in Section~\ref{ch:traindata}. Fig.~\ref{fig:mibench-9} shows the measurement results of the Mibench benchmarks. 

The testing data collected will be used to evaluate the software estimation models developed in the next subsection.

In general, the measurements of the utilization of these 9 benchmarks are more stable and do not vary much.

\begin{figure}[h]
  \centering
  \includegraphics[width=0.9\linewidth]{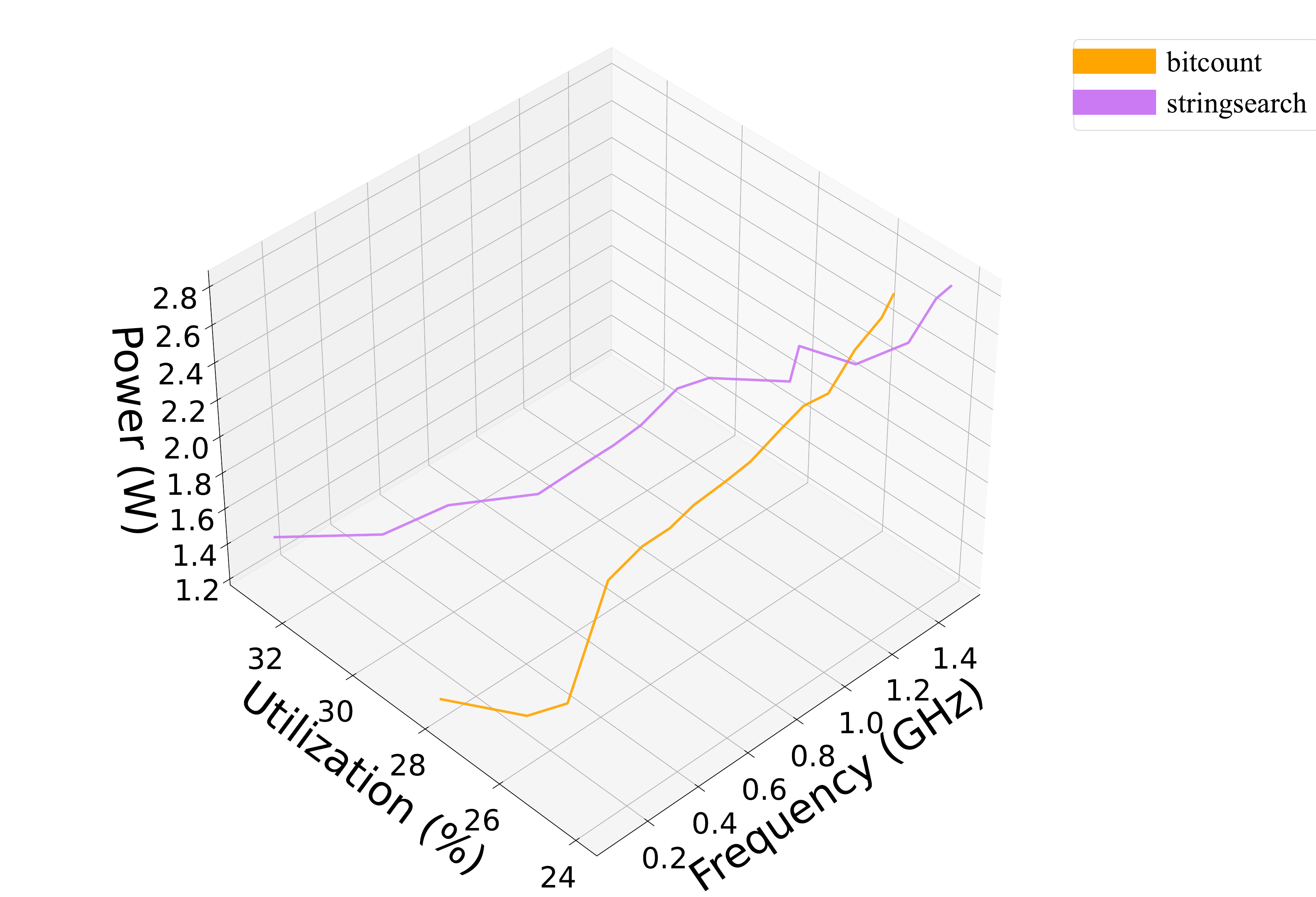}
  \caption{$stringsearch$ vs $bitcount$ }
  \label{fig:mibench-2}
\end{figure}

The $stringsearch$ benchmark in Mibench has a different trend than the other nine benchmarks.
In Fig.~\ref{fig:mibench-2}, one of the nine benchmarks, $bitcount$,  is used to show the difference with $stringsearch$.   It can be seen from the figure that the power of $bitcount$  has a jump (a similar trend can be found in the 9 benchmarks other than $stringsearch$ in Fig. \ref{fig:mibench-9}). On the other hand, the power of $stringsearch$ increases progressively with increasing frequency. We can observe that the nine benchmarks show little change in utilization at all frequencies while  $stringsearch$ has higher utilization at lower frequencies than the other benchmarks. At high frequencies, $stringsearch$'s utilization is reduced. Therefore no jump in energy consumption of this benchmark.

\begin{figure}[h]
  \centering
  \includegraphics[width=0.9\linewidth]{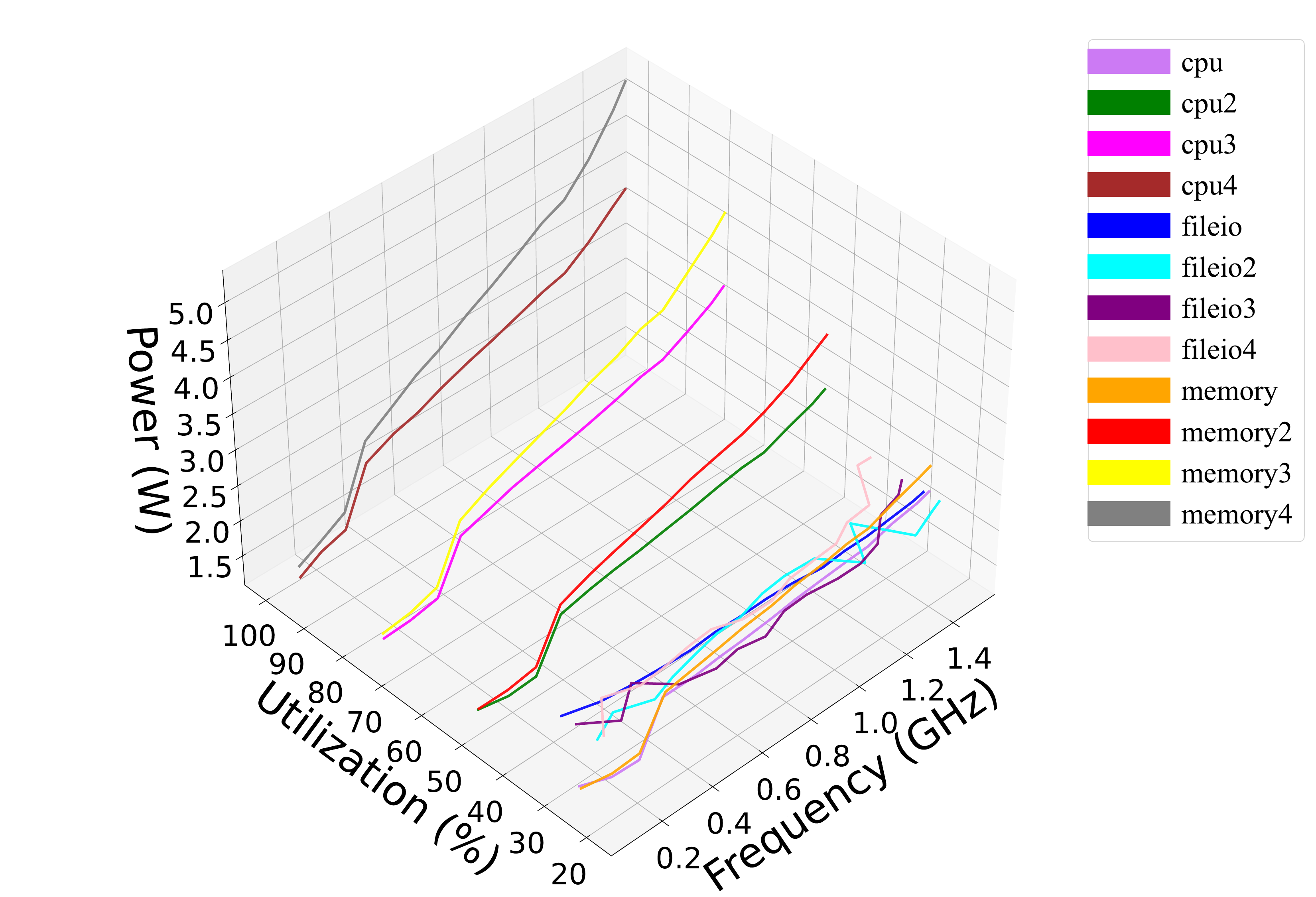}
  \caption{Measurement results of Sysbench}
  \label{fig:Sysbench}
\end{figure}

The measurement results of Sysbench are shown in Fig.~\ref{fig:Sysbench}. 
It can be seen from the figure that for $cpu$ and $memory$, their thread count is reflected in the variation of utilization, which is 30\% utilization for a single thread, 60\% utilization for 2 threads, 80\% utilization for 3 threads and 100\% utilization for 4 threads. Their power also appears as a jump with frequency.

\subsection{Model Derivation and Evaluation}
\label{sec:ModelDesignEval}

We use the data collected (illustrated in Section~\ref{ch:resultsoftraindata}) to build the models. \textcolor{black}{We have explored linear regression-based, decision-tree-based and neural-networked-based software models for power estimation, which will be described in detail in the later part of this section. }

\textcolor{black}{We use the benchmarks in Table~\ref{table:benchmarks} }  %from Section~\ref{ch:testdata}
to evaluate the prediction accuracy of the models. The metrics used are mean squared error (MSE), mean absolute error (MAE), and $R^2$ score \cite{ash1999r2}. \textcolor{black}{Mean Squared Error (MSE) is the average of the squares of the errors—that is, the average squared difference between the estimated values and the actual value. Mean Absolute Error (MAE) is the average of the absolute errors—the average absolute difference between the estimated values and the actual value. The $R^2$ score indicates the degree of explanation of the dependent variable by the independent variable. It ranges from 0 to 1. The higher the value of this indicator, the higher the degree of prediction accuracy.}

\begin{algorithm}
\caption{\textcolor{black}{Model Evaluation for Model $M$}}
\label{algo:Evaluation}
\begin{algorithmic}[1]

    \State \textbf{Set} $governor$ to \texttt{userspace}
    \State \textbf{Initialize} $MSE \gets 0$
    \State \textbf{Initialize} $MAE \gets 0$
    \State \textbf{Initialize} $n \gets 0$
    \State \textbf{Initialize} $SumP \gets 0$
    \State \textbf{Initialize} empty lists $P\_list$ and $P^M\_list$
    \For{each benchmark program $i$}
        \For{each frequency $j$ that the device supports}
            \State \textbf{Run} benchmark $i$ at frequency $j$
            \State \textbf{Measure} utilization $u_{ij}$ and actual power $P_{ij}$
            \State \textbf{Calculate} predicted power $P^M_{ij}$ using model $M$ with $u_{ij}$ and frequency $j$
            \State $MSE \gets MSE + (P_{ij} - P^M_{ij})^2$
            \State $MAE \gets MAE + |P_{ij} - P^M_{ij}|$
            \State \textbf{Append} $P_{ij}$ to $P\_list$
            \State \textbf{Append} $P^M_{ij}$ to $P^M\_list$
            \State $SumP \gets SumP + P_{ij}$
            \State $n \gets n + 1$
        \EndFor
    \EndFor
    \State \textbf{Compute} mean actual power $\overline{P} = \dfrac{SumP}{n}$
    \State \textbf{Compute} $MSE \gets \dfrac{MSE}{n}$
    \State \textbf{Compute} $MAE \gets \dfrac{MAE}{n}$
    \State \textbf{Initialize} $SSR \gets 0$ \Comment{Sum of Squares of Residuals}
    \State \textbf{Initialize} $SST \gets 0$ \Comment{Total Sum of Squares}
    \For{$k \gets 1$ to $n$}
        \State $SSR \gets SSR + (P\_list[k] - P^M\_list[k])^2$
        \State $SST \gets SST + (P\_list[k] - \overline{P})^2$
    \EndFor
    \State \textbf{Compute} coefficient of determination $R^2 = 1 - \dfrac{SSR}{SST}$
    
\end{algorithmic}
\end{algorithm}

\textcolor{black}{
How to evaluate  a given power estimation Model $M$ with the above metrics is shown in Algorithm~\ref{algo:Evaluation}. To obtain the measured values for comparison with the predicted values, we configured each benchmark to run at various frequency settings and measured the corresponding utilization and power consumption. Specifically, we recorded each measurement in the format of benchmark identifier, utilization, frequency, and power. Given that there are multiple frequency settings (referred to as frequency steps) and several different benchmarks, this process yielded a total number of data points ($n$ in Algorithm~\ref{algo:Evaluation}) equal to the product of the number of frequencies and benchmarks. After collecting the actual power consumption data, we applied the models discussed in the following sections to predict the power values. We then compared the actual power data with the predicted power data to evaluate the performance indicators.}

Next, we describe the software power estimation models explored on the Jetson device.
\subsubsection{Polynomial Regression Model}
\label{ch:regression}

We constructed four linear models based on the training data. 
The first model (Model 1) we choose is the simplest polynomial model. We call Model 1 the Simple Regression Model. Frequency and utilization as two parameters are selected in the model as cubic and primary, respectively. This is because, in the CMOS circuit model, frequency and power consumption relation are cubic.
The second model is optimized based on Model 1. It added more terms with different exponents, with the highest order being 3. The reason for this modification is simply the intuitive idea that adding more terms will improve the accuracy of the model. We call it Model 2, the Multi-Term Regression Model.
Considering the frequency/voltage levels, we further divide Model 2 into two different equations corresponding to 2 frequency/voltage levels. It uses one set of coefficients when the frequency is in the first 3 levels, and another set of coefficients when the frequency is in the later levels. We call it the Multi-Frequency Regression Model (Model 3).
 We improve Model 3 further and introduce Model 4, called the Per-Frequency Regression Model, which fits data to individual frequency.

\textcolor{black}{Based on the test data from ~\ref{ch:testdata}, we evaluated the performance of the four models and the results are summarized in the Table}~\ref{tab:linearmodel}.

\begin{table}[htbp]
  \centering
  \caption{The accuracy of regression models}
  \label{tab:linearmodel}
    \begin{tabular}{|c|c|c|c|c|}
        \hline
        &
        \textbf{Model} & \textbf{MSE} & \textbf{MAE} & \textbf{$R^2$ score} \\
        \hline 
        1 &
        \textbf{Simple Regression Model} & 0.2397 & 0.4184 & -0.0242 \\
        \hline
        2 & 
        \textbf{Multi-Term Regression Model} & 0.0272 & 0.1365 & 0.8839 \\
        \hline
        3 &
        \textbf{Multi-Frequency Regression Model} & 0.0270 & 0.1344 & 0.8848 \\
        \hline
        4 &
        \textbf{Per-Frequency Regression Model} & 0.0182 & 0.1040 & 0.9221 \\
        \hline
    \end{tabular}
\end{table}

The results show that Per-Frequency Regression Model performs much better than the previous three models, with the lowest MSE and MAE and the highest $R^2$ score.

\textcolor{black}{An $R^2$ score of 0.9221 implies that approximately 92.21\% of the variance in power consumption can be explained by the model using frequency and utilization as predictors. This high $R^2$ value indicates a strong correlation between the predicted and actual power consumption values, signifying that the model fits the data well and has high predictive power.}

\textcolor{black}{The high $R^2$ score of the Per-Frequency Regression Model indicates that it can reliably predict the power consumption of the Jetson Nano Board (2GB) under various workloads and operating conditions. }

\subsubsection{Decision Tree Model}
%  HYW add the definition of MSE MAE.
The following two models use the decision tree model. The accuracy of 2 decision tree models is shown in Table~\ref{tab:treemodel}.

\begin{table}[htbp]
  \centering
  \caption{The accuracy of decision tree models}
  \label{tab:treemodel}
    \begin{tabular}{|c|c|c|c|}
        \hline
        \textbf{Model} & \textbf{MSE} & \textbf{MAE} & \textbf{$R^2$ score} \\
        \hline
        \textbf{Decision Tree Model 1} & 0.1361 & 0.3052 & 0.4187 \\
        \hline
        \textbf{Decision Tree Model 2} & 0.0433 & 0.1668 & 0.8149 \\
        \hline
    \end{tabular}
\end{table}

Decision tree model 1 is a model fitted by a machine learning algorithm based on XGBoost \cite{chen2016xgboost}. It trains and integrates decision tree models using gradient-boosting methods and generates a forest for making a decision. 

The results show that this model performs worse than the regression models 2-4, with an MSE of 0.1361 and MAE of 0.3052. Its $R^2$ score of 41.87\%, which is low. 
%indicates it explains 41.87\% of the variance in the data.
Obviously, this model does not work well.

Decision tree model 2  enhances Decision tree model 1 by applying the XGBoost algorithm for each frequency, using a similar idea in the construction of polynomial regression model 4. 
%It is constructed by the same algorithm with the same inputs and outputs as decision tree model 1. 
The results show that Decision Tree Model 2's performance is much better than Decision Tree Model 1, with a lower mean squared error of 0.0433 and a slightly lower mean absolute error of 0.1668. Its $R^2$ score of 0.894. %indicates it explains approximately 81.49\% of the variance in the data.

\subsubsection{Neural Network Model}
The above six models constructed are based on explicit parameters: frequency and average utilization. 
We also tried to construct models with more input data. We replace the original average utilization with the utilization per core. Because of the more input parameters used, we train it with a neural network model.

We use the same training data as before, that is measurements from the multi-threaded CPU-intensive benchmark. The utilization of each core has been recorded in the previous measurements. The defined neural network has 5 inputs and 1 output. The 5 inputs correspond to the frequency and the utilization of the 4 cores and the 1 output is the predicted power consumption. The middle layer consists of a total of 4 hidden layers with 128, 64, 32 and 16 neurons. 

The Neural Network model is trained 4000 times with the training data.  
The model was then tested with the testing data. The MSE is 0.0328, MAE is 0.1346. This model is not as effective as the best regression model 4, but it is still a plausible model.

\section{Power Prediction for Programs running in Real-World Environment}
\textcolor{black} {An operating system that supports power governors can regulate frequency to save energy. Thus, a program is unlikely to run with only one frequency.  Predicting the power of programs running with variable frequencies is important. This section introduces our method for predicting the power of programs running in a real environment using the device's derived power model together with a data tracker software module.
}

\subsection{Kernel-Level Data-Tracker Module}
\label{ch:module}
\textcolor{black} {We have developed a Kernel-level Data Tracker module to monitor real-time CPU frequency changes and log the corresponding CPU load data. The Kernel-level Data Tracker, together with the derived power model,  is essential for software-based energy prediction. This module leverages the Linux $cpufreq$ API, specifically utilizing the $get\_cpu\_idle\_time$ function, which retrieves the CPU's idle time over a period. By tracking these changes, the module provides data for power estimation and performance analysis.}

Each CPU in the system has a dedicated $cpu\_data$ structure that stores the previous idle and total times of the CPU. This structure is crucial for calculating the CPU load, as it allows the module to determine the difference in idle and total times between frequency changes.

\subsubsection{Core Mechanism}

When the CPU frequency changes, a callback function $frequency\_notifier$  is triggered, which responsible for recording the old frequency, the duration for which it was maintained, and the CPU load during this period. The following steps outline the process:

\begin{itemize}
    \item \textbf{Frequency Change Detection}: The module registers a notifier for CPU frequency transitions using the $cpufreq\_register\_notifier$ function. This notifier is called whenever the CPU frequency changes.
    \item \textbf{Data Logging}: The notifier function logs the old frequency, the duration it was active, and the load of each CPU. The data is formatted into a string and written to a specified file path \textit{data\_file\_path} using Linux file system APIs.
    \item \textbf{CPU Load Calculation}: The CPU load is calculated by comparing the current idle and total times with the previously recorded values. This calculation is performed using the $cpu\_load$ function, which ensures that negative values are handled appropriately and that the load percentage does not exceed 100\%.
\end{itemize}

\subsubsection{Data Persistence}

To persist the tracked data, the module writes the information to a file specified by \textit{data\_file\_path}. The file operations are performed in kernel space using the $vfs\_write$ function. The data is appended to the file, ensuring that all frequency changes and corresponding loads are recorded sequentially.

\subsubsection{Proc File System Interface}

The module provides an interface through the $proc$ file system, allowing users to control the logging functionality. By writing specific symbols ('1' to start logging, '0' to stop logging) to the $cpufreq\_monitor$ proc file, users can enable or disable the logging process dynamically. This interface is implemented in the $procfile\_write$ function, which interprets user input and updates the logging state accordingly.

\subsubsection{Stability and Error Handling}

To enhance the stability and reliability of the module, several error handling mechanisms are included:

\begin{itemize}
    \item \textbf{Negative Time Handling}: The $cpu\_load$ function checks for negative values in idle and total time calculations. If such values are detected, a warning is logged, and appropriate measures are taken to ensure the calculations remain valid.
    \item \textbf{Load Percentage Bounds}: The module ensures that the calculated CPU load does not exceed 100\%. If an anomaly is detected, such as a load percentage greater than 100\%, a warning is logged for further investigation.
    \item \textbf{File Operation Errors}: The $log\_data$ function includes checks to handle errors that may occur during file operations, such as failure to open the file. Appropriate error messages are logged to help diagnose issues.
\end{itemize}

\subsubsection{Portability}

The module is built using standard Linux kernel APIs, making it highly portable across different hardware platforms. It can be deployed on x86, ARM, or any other Linux-enabled architecture without requiring hardware-specific modifications. This portability ensures that the module can be used in a wide range of environments, from desktop systems to embedded devices.

%This module offers a robust solution for monitoring CPU performance, enhancing the understanding of power consumption and enabling optimized resource management across diverse computing environments.
The Data Tracker module's design and implementation %significantly 
contribute to performance analysis, provide a reliable and versatile mechanism for real-time CPU monitoring, enhance the understanding of power consumption, and enable optimized resource management across diverse computing environments.

\subsection{Example: Power Prediction for programs with Multiple Frequencies}

\begin{figure*}[h]
  \centering
  \includegraphics[width=\textwidth]{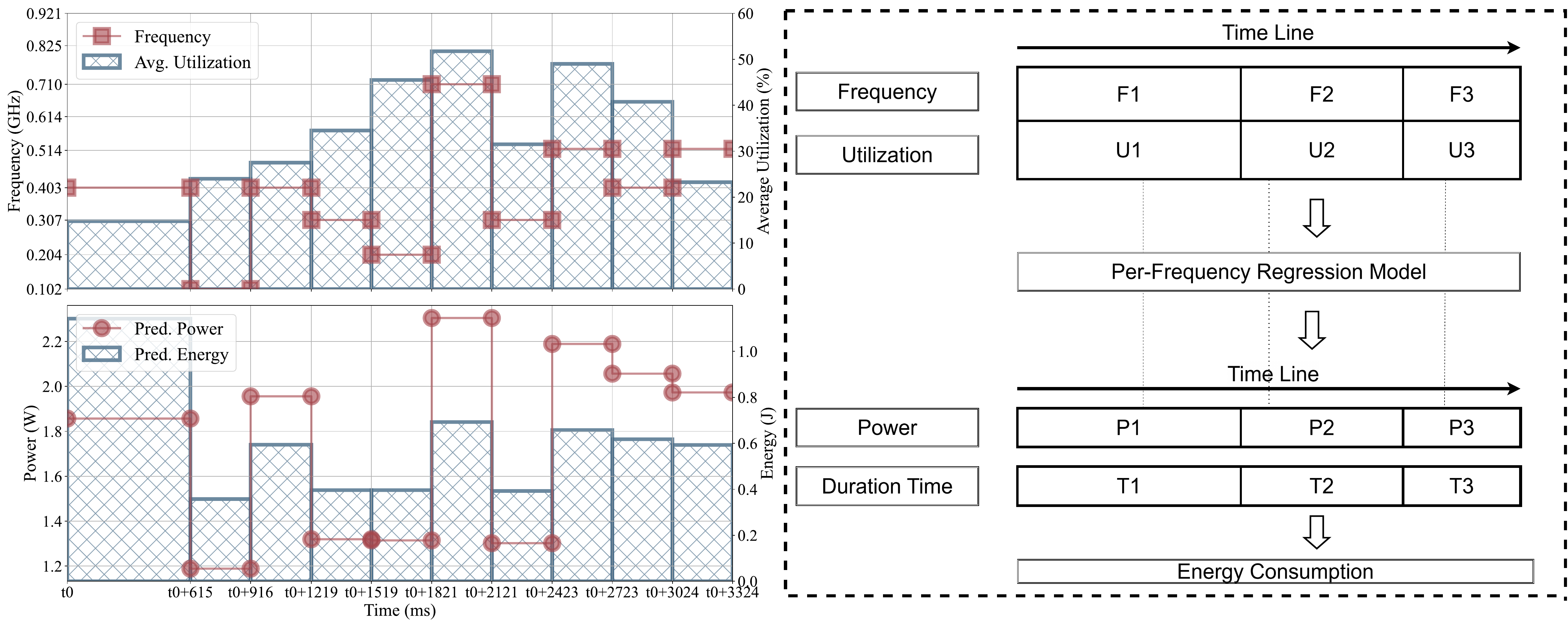}
  \caption{\textcolor{black}{Actual trace of utilization and frequency with predicted power consumption}
  }
  
  \label{fig:ondemad_explanation}
\end{figure*}

%{Figure~\ref{fig:ondemad}
Fig.~\ref{fig:ondemad_explanation} shows a portion of the real trace derived from the Data Tracker module. The trace includes the utilization and frequency of a workload (\textcolor{black}{top-left}) and the predicted power values and Energy Consumption (\textcolor{black}{bottom-left}) for each frequency segment. 
The trace is the experimental result for the benchmark $basicmath$ running with variable frequencies under the $ondemand$ governor. The x-axis represents time, while the y-axis conveys information about each frequency segment. In the upper part of Fig.~\ref{fig:ondemad_explanation}, the height of the \textcolor{black}{striped blue bar} indicates the frequency level, and the height of the \textcolor{black}{red square markers and lines} denotes the utilization at that frequency. The lower part of Fig.~\ref{fig:ondemad_explanation} shows the predicted power values (depicted by the \textcolor{black}{red circular markers and lines}) for each frequency segment, calculated using our constructed power model, and the energy consumption calculated by duration times power (\textcolor{black}{depicted by the height of the striped blue bar}).  \textcolor{black}{The right part of Fig.~\ref{fig:ondemad_explanation} depicts how the power of each segment is calculated (top to bottom) and the calculation across time (left to right).}

%\begin{figure*}[h]
%  \centering
  %\includegraphics[width=\linewidth]{figure/od_v1.png}
  %\caption{Frequency, utilization and predicted power of $basicmath$ run under $Ondemand$ }
  %\label{fig:ondemad}
%\end{figure*}

\textcolor{black}{%The figure only shows a portion of the real trace of the workload. 
The predicted power value based on the entire trace of collected data is 2.137. The measured true value is 2.285. The power prediction model used is the Per-frequency Regression Model (Model 4). % which is identified as the optimal model in Section~\ref{sec:ModelDesignEval}.  
The low prediction error demonstrates that our model is applicable for power prediction under multiple frequencies with high accuracy.}

\section{Evaluation of Power Prediction and Discussion}

In this section, we  evaluate our models by comparing the predicted results with empirical data for programs executed under variable frequencies and discuss the implications of our findings. The evaluation includes testing programs that runs under the Linux $ondemand$ governor, which dynamically adjusts frequencies based on system utilization. We also  deploy our approach on a different embedded platform to validate its reproducibility.

\subsection{Evaluation Under $ondemand$ Governor}

We tested various benchmarks running under the Linux $ondemand$ governor to assess the accuracy of our power prediction model. The $ondemand$ governor dynamically adjusts the CPU frequency based on system utilization, which necessitates capturing all frequencies, utilizations, and durations for which the underlying program is executed. We achieved this by incorporating the data tracking module previously discussed in Section~\ref{ch:module}, which records data each time a frequency change occurs.

\begin{figure}[h]
  \centering
  \includegraphics[width=\linewidth]{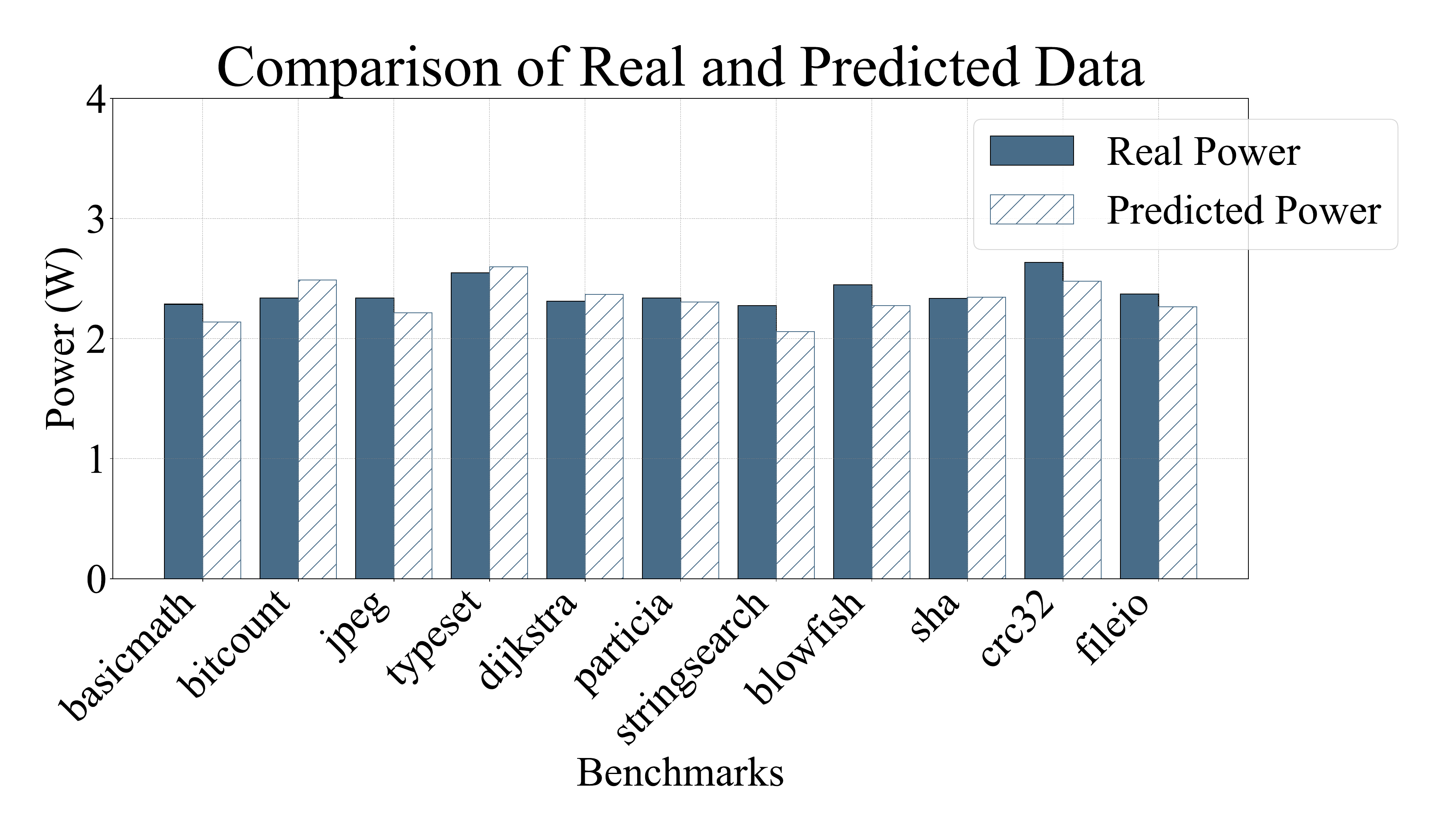}
  \caption{\textcolor{black}{Benchmark predictions vs. real values on JTN}}
  \label{fig:ondemand_all}
\end{figure}

Fig.~\ref{fig:ondemand_all} shows the comparison between the measured data and the predicted data for 11 benchmarks. The \textcolor{black}{dark blue} bars represent the measured values, while the \textcolor{black}{striped light blue} bars indicate the values predicted by our model. As observed, the predicted and actual values are closely aligned, signifying the high prediction accuracy of our model.

Table~\ref{tab:absolute} presents the absolute error for each benchmark's prediction. The mean absolute error across all benchmarks is 0.1118, and the mean squared error is 0.0164. These metrics confirm that the prediction error is minimal, thereby validating the high prediction accuracy of our model.

\begin{table}[htbp]
  \centering
  \caption{Absolute error}
  \label{tab:absolute}
    \begin{tabular}{|c|c||c|c|}
        \hline
        \textbf{Benchmark} & \textbf{AE} & \textbf{Benchmark} & \textbf{AE} \\
        \hline
        $basicmath$ & 0.148 & $bitcount$ & 0.1506 \\
        \hline
        $jpeg$ & 0.1236 & $typeset$ & 0.0499 \\
        \hline
        $dijkstra$ & 0.0576 & $patricia$ & 0.0341 \\
        \hline
        $stringsearch$ & 0.2163 & $blowfish$ & 0.1739 \\
        \hline
        $sha$ & 0.0091 & $crc32$ & 0.1592 \\
        \hline
        $fileio$ & 0.1063 \\
        \cline{1-2}
    \end{tabular}
\end{table}

\subsection{Reproducibility on Different Platforms}

To further validate our approach, we also deployed it on a different embedded platform, specifically, the Raspberry Pi. The validation process includes data collection, model building, prediction of simulated behavioural data, and prediction of real behavioural data.

\begin{figure}[h]
  \centering
  \includegraphics[width=\linewidth]{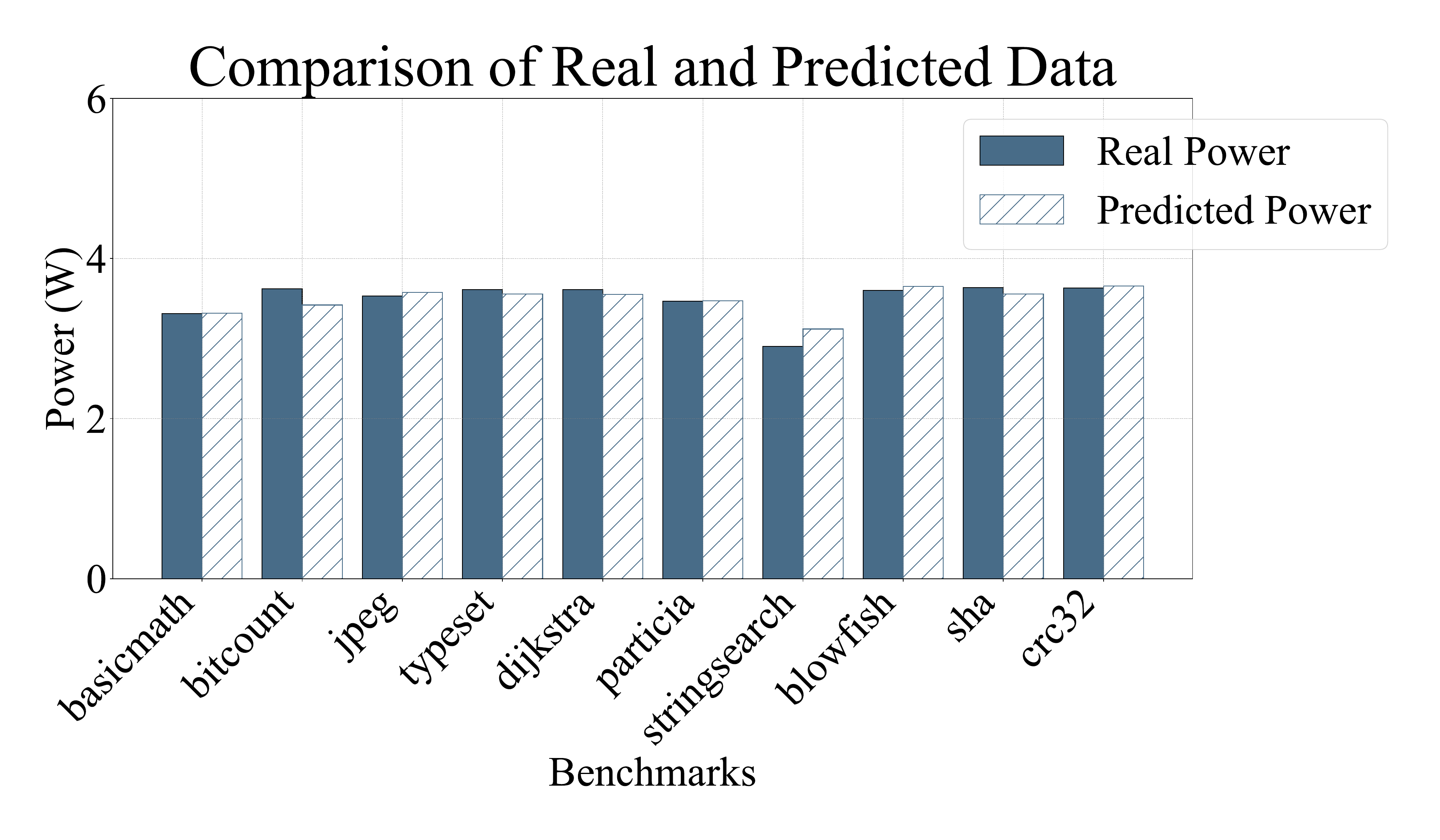}
  \caption{\textcolor{black}{Benchmark predictions vs. real values on RBP}}
  \label{fig:ondemand_all_rbp}
\end{figure}

Fig.~\ref{fig:ondemand_all_rbp} presents the benchmark predictions versus real values on the Raspberry Pi. The results indicate that the average mean error is approximately 4\%, with the best test achieving an average error of 2\%. These results demonstrate that the methodology is applicable to different embedded platforms for achieving accurate energy consumption predictions.

\textcolor{black}{However, certain discrepancies were observed. For instance, the $fileio$ benchmark behaves differently on the Raspberry Pi. This could be due to the specific behavior of file I/O operations on this platform. }
%Additionally, benchmarks with higher utilization
%rates tend to have poorer prediction accuracy, likely due to
%differences in power consumption between programs written
%in different languages

%Additionally, benchmarks with higher utilization rates tend to have poorer prediction accuracy on Raspberry Pi. Our further studies show that the power consumption for programs written in different languages (Python and C) are different.}

\textcolor{black}{\subsection{Limitations}}

\textcolor{black}{While our proposed method demonstrates high accuracy in power estimation for devices like the Jetson Nano and Raspberry Pi, it has certain limitations.}

\textcolor{black}{Our power estimation model focuses on applications with primarily CPU-Centric power consumption. That is, the CPU is the primary energy consumer in embedded devices. For applications running on devices where other components such as GPUs, high-speed memory (e.g., LPDDR4 RAM), or solid-state drives (SSDs) have significant power consumption, our model may not accurately capture the total energy usage. For instance, devices performing intensive graphics processing with GPUs, conducting high-throughput memory operations, or engaging in frequent disk I/O with SSDs may exhibit power profiles our CPU-focused model does not account for, leading to reduced overall accuracy in power estimation for such scenarios.}

% Future work could involve extending the model to incorporate power consumption from other significant components. This may include monitoring GPU utilization, memory bandwidth usage, and storage I/O activity to enhance the model's comprehensiveness and improve estimation accuracy in scenarios where these components contribute substantially to power consumption.

\textcolor{black}{
%Applicability to Homogeneous Architectures:
Our method has been tested on devices with homogeneous multi-core CPUs, such as the Jetson Nano and Raspberry Pi, which consist of identical cores. The applicability of our method to heterogeneous systems with different types of cores, such as big.LITTLE architectures commonly found in modern mobile processors, has not been evaluated. In these systems, cores may have different performance and power characteristics, making power estimation more complex.}

% Additional considerations and adaptations of the model may be necessary to accurately estimate power consumption in heterogeneous architectures. For example, separate power models might be required for the "big" and "LITTLE" cores, accounting for their distinct characteristics. Exploring this extension could be a direction for future research 

Overall, our model demonstrates high prediction accuracy under varying conditions and on different platforms, validating its robustness and applicability.

\section{conclusion}
\textcolor{black}{
Most small inexpensive embedded devices in MEC systems do not have software-based power measurement support.
This paper proposes a smart data-driven method using consumer-grade meters to construct software-level power estimation models for such embedded devices.} 
\textcolor{black}{We first describe the  drawbacks associated with the use of a power meter to measure energy consumption, which are the inability to measure instantaneous power consumption and the inability to act on energy feedback-based systems.} Then, we describe our approach to constructing the power model by systematically generating workloads and collecting training data.
To use our approach, one can follow the steps described in this paper to build the model for their own devices. The steps include building the workload, collecting training data, constructing the model and using the model.
During the experiments, we explored several types of models, and in the end, the model that worked best on the device that we tested was a per-frequency polynomial regression model on utilization. The model accuracy is 92\%  compared to the long-duration measurement. \textcolor{black}{The approach is straightforward and systematic, capable of achieving good results in predicting energy consumption. This allows the researchers/application developers to eliminate the need of using a power meter for power measurement once the model is established.} We also provide a platform-independent kernel module that can be used to collect running traces of utilization and frequencies needed for power prediction for programs running in Linux environment.

\bibliographystyle{IEEEtran}
\bibliography{reference}

\begin{IEEEbiographynophoto}{Haoyu Wang}
Haoyu Wang received a B.S. degree in Computer Science from St. Francis Xavier University, Canada. He is currently a graduate student in the Faculty of Computer Science at Dalhousie University, Canada. His research interests include Linux kernel optimization, DVFS, and energy performance trade-off in embedded systems.
\end{IEEEbiographynophoto}

\begin{IEEEbiographynophoto}{Xinyi Li} 
Xinyi Li received her Bachelor's degree in Computer Science from St. Francis Xavier University, Antigonish, Canada. She is currently pursuing a Master's degree at Dalhousie University, Halifax, Canada. Her research focuses on cybersecurity and machine learning.
\end{IEEEbiographynophoto}

\begin{IEEEbiographynophoto}{Ti Zhou}
Ti Zhou received his Bachelor’s and Master’s degrees in Computer Science from St. Francis Xavier University, Antigonish, Canada. He is currently pursuing a Ph.D. in Computer Science at Stony Brook University, USA. He has a broad interest in computer systems, with a current focus on verification of distributed systems and energy efficiency.
\end{IEEEbiographynophoto}

\begin{IEEEbiographynophoto}{Man Lin}
Man Lin received her B.E. degree in Computer Science and Technology from Tsinghua University, China, in 1994. She received the Lic. and Ph.D. degrees from the Department of Computer Science and Information at Linkoping University, Sweden, in 1997 and 2000, respectively. She is currently a Professor of Computer Science at St. Francis Xavier University, Canada. Her research interests include real-time and cyber-physical system design and analysis, scheduling, power-aware computing, and optimization algorithms. Her research is supported by the National Science and Engineering Research Council of Canada (NSERC). 
\end{IEEEbiographynophoto}

\end{document}